# Calibration of the Makrofol Nuclear Track Detector using Relativistic Lead Ions

S. Cecchini*, T. Chiarusi*, M. Cozzi*, D. Di Ferdinando*, M. Frutti*, G. Giacomelli*,
A. Kumar*,2, G. Mandrioli*, S. Manzoor*,1, E. Medinaceli*,3, L. Patrizii*, V. Popa*,4,
M. Spurio*, V. Togo* and I.E. Qureshi[1]

* *University of Bologna and INFN Bologna, V. Berti-Pichat 6/2, I-40127 Bologna, Italy*
1  *Physics Research Division, PINSTECH, P.O. Nilore Islamabad, Pakistan*
2  *Dept. of Physics, SHSL-Central Institute of Engg. & Tech., Longowal, India*
3  *Inst. Invest. Fisicas, La Paz, Bolivia*
4  *ISS, R-77125, Bucharest, Romania*

**Abstract**

We present the calibration of the Makrofol nuclear track detector (NTD) using Pb-ions of 158 GeV/amu. Improvements of the post etched surfaces, reduction in the surface background of the Makrofol NTDs and high contrast tracks were achieved with the appropriate addition of ethyl alcohol in KOH aqueous solutions. The calibration of Makrofol has shown for the first time all the peaks due to nuclear fragments with $Z \geq 52$. The measurement of the cone heights shows well-separated individual peaks for $Z = 59 - 82$ and 83 (charge pickup).

Keywords: SLIM experiment; chemical etching, REL, magnetic monopoles, relativistic nuclei.

___________
*Corresponding author. Tel.: +39-051-2095235; fax: +39-051-2095269
E-mail address: manzoor@bo.infn.it

## 1. Introduction

Plastic nuclear track detectors (NTDs) are used in the searches for magnetic monopoles, nuclearites and nuclear fragments with fractional charges, for the determination of the fragmentation cross-sections, for measurement of the composition of primary cosmic rays, etc (Ambrosio et al., 2002; Cecchini et al., 1993; Chiarusi et al., 2003, Kumar et al., 2003).

The main aim of the present work is the determination of the optimal etching conditions of the Makrofol NTDs used in the SLIM experiment as well as in other experiments (Cecchini et al., 2001). Previously we used aqueous solutions of NaOH and KOH (Cecchini et al., 1996; Giacomelli et al., 1997; Manzoor et al., 2000). The addition of ethyl alcohol in the etchant improves the etched surface quality and reduces the number of surface defects and background tracks.

The Makrofol polycarbonate was manufactured by Bayer, Germany. For the calibration of Makrofol detectors we used foils of 11.5 cm x 11.5 cm, 500 μm thick and we exposed them in 1996 to the 158 GeV/amu Pb-ions at the CERN-SPS. The etchants used were water solutions of 6N KOH at different temperatures (50° to 65°C), and with different percentages of ethyl alcohol.

Experimental details are given in Section 2. The calibrations are discussed in Section 3 and the conclusions in Section 4.

## 2. Experimental details

A stack composed of Makrofol and CR39 foils with a 10 mm thick aluminium target was exposed to 158 GeV/amu Pb-ions at the CERN-SPS. The detector foils downstream of the target recorded the survived Pb-ions as well as their fragments. After exposure, two Makrofol foils were etched in 6N KOH + 20% (by volume) alcohol for 8 hours. After etching, the cone base areas were measured with an Elbek automatic image analysis system (Noll et al., 1988).

It was observed that the bulk etch rate increases with the increase of the etching temperature and the percentage of the alcohol mixture in the etchant, as shown in Table 1. The presence of the alcohol also polishes the detector surface and improves considerably the sharpness of the post-etched surface of the detector. With the etching condition N.2 in table 1, we obtained tracks quite suitable for the ELBEK automatic measuring system. The bases of the cones are more regular and the transparency of the detector is much better as compared to the etching without ethyl alcohol. A special effort was made to keep the etching tanks tight and keep a good stability of the etching conditions.

Table 1: Bulk etching rates ($v_B$) for Makrofol NTDs at different etching temperatures.

| S.N | Etching Conditions | $v_B$ (μm/h) |
|---|---|---|
| 1 | 6N NaOH, 50 °C | 0.24 ± 0.01 |
| 2 | 6N KOH + Ethyl alcohol, 50 °C  80 : 20 % by volume | 2.97 ± 0.05 |
| 3. | 6N KOH + Ethyl alcohol, 50 °C  50 : 50 % by volume | 5.87 ± 0.05 |
| 4. | 6N KOH+ Ethyl alcohol, 60 °C  80 : 20 % by volume | 13.0 ± 1 |
| 5. | 6N KOH+ Ethyl alcohol, 65 °C  80 : 20 % by volume | 23.0 ± 1 |

## 3. Calibration

Figure 1 shows some tracks of 158 GeV/amu Pb-ions and of their fragments in Makrofol, etched in NaOH and KOH solutions with and without alcohol. Notice the improvements in surface quality and the background reduction with the addition of ethyl alcohol in the etching solution.

Figure 2 shows the average base area distribution for Pb-ions and their fragments in Makrofol; averages were made over measurements on two front sides. Each peak of the distribution in Fig. 2 corresponds to a nuclear fragment of different charge.

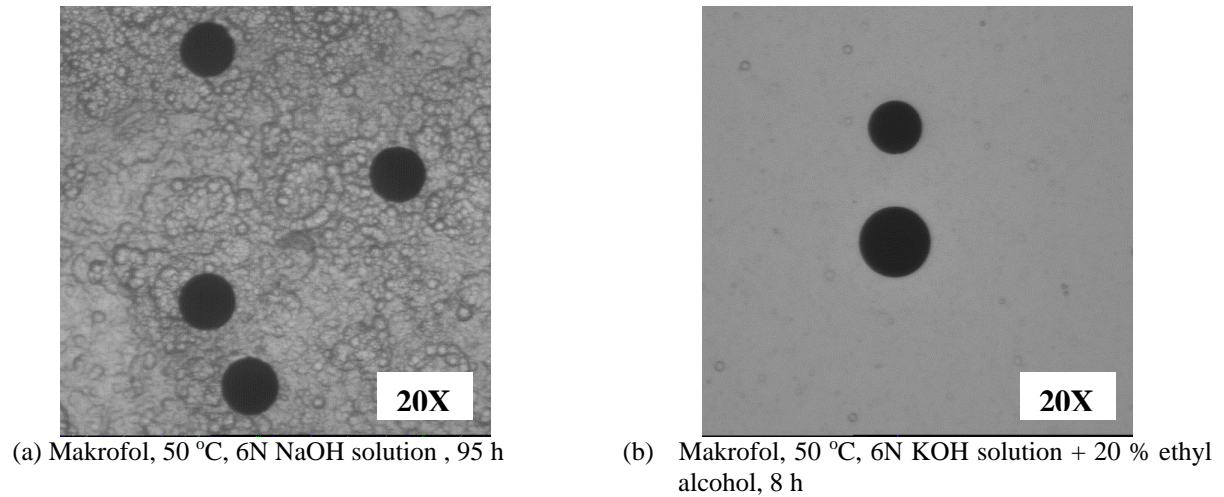

(a) Makrofol, 50 $^{o}$C, 6N NaOH solution , 95 h    (b) Makrofol, 50 $^{o}$C, 6N KOH solution + 20 % ethyl alcohol, 8 h

Figure 1. "Tracks" of Pb-ions and their fragments in Makrofol nuclear track detectors under different etching conditions.

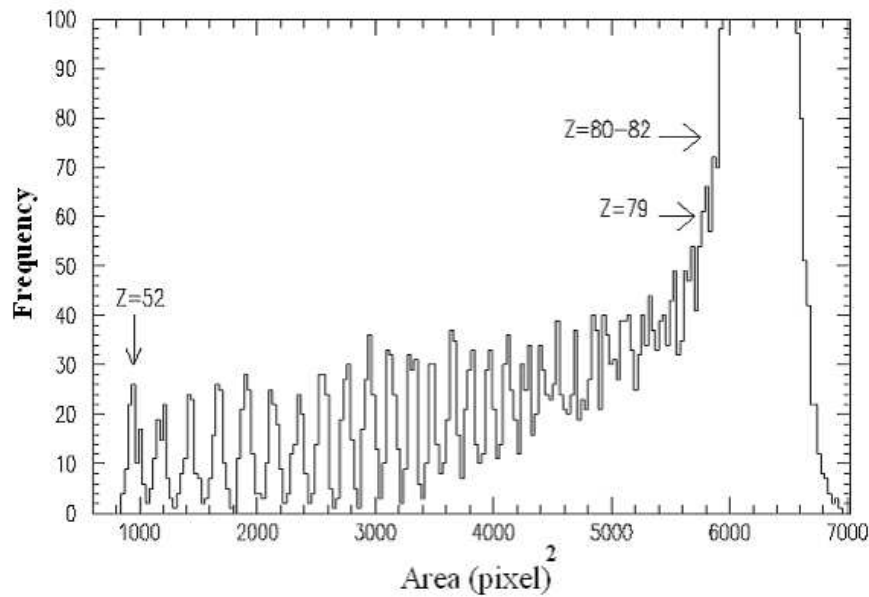

Figure 2. Distributions of the base areas of the etch-pit cones from relativistic Pb-ions in Makrofol.

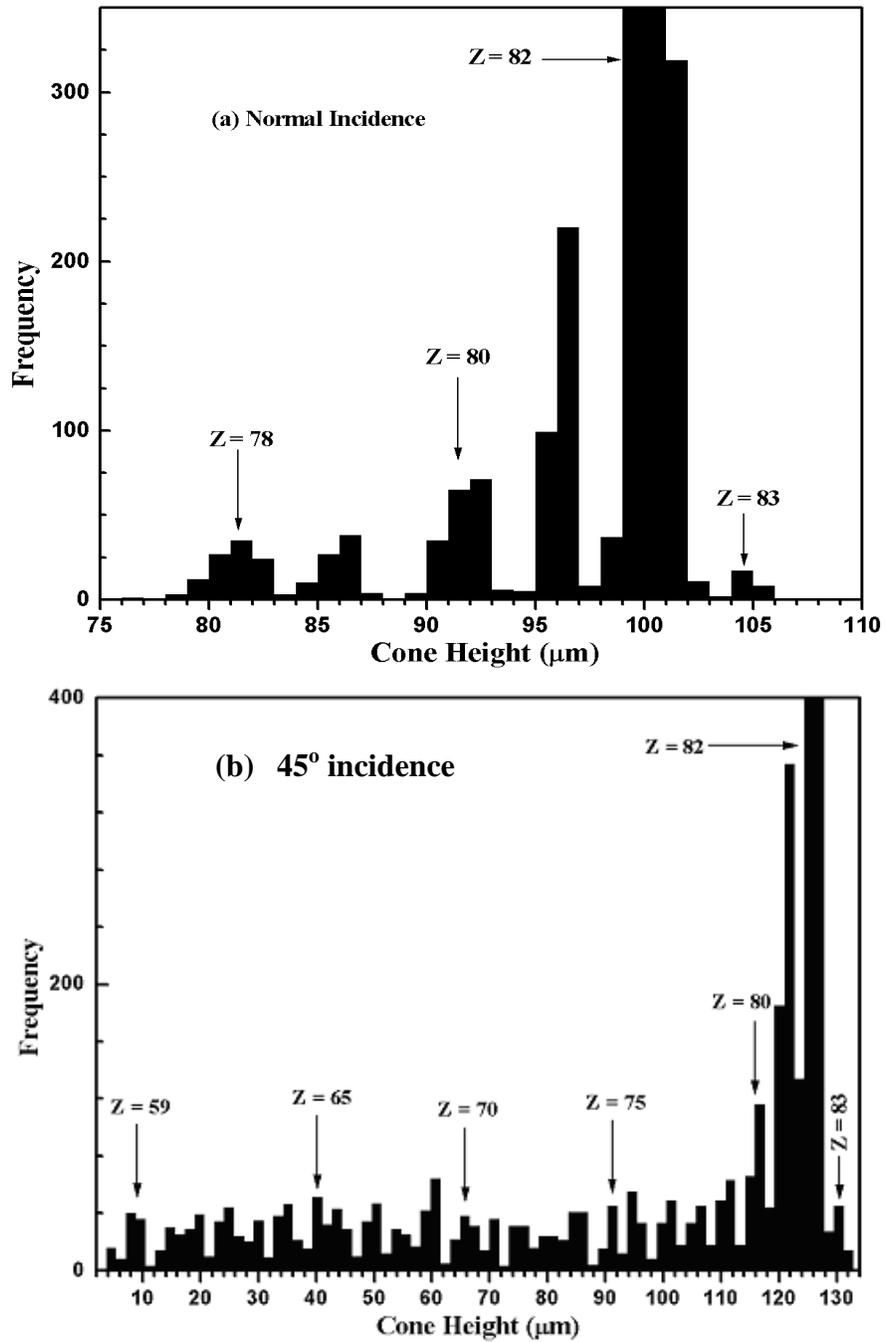

Figure 3. Distributions of the cone heights in Makrofol detectors exposed to relativistic Pb-ions at (a) normal incidence and (b) 45° incidence.

The charge resolution of peaks close to the Pb peak (Z = 82) can be improved by measuring the heights of the etch-pit cones. Two Makrofol foils exposed one at 90° and the other one at 45° with respect to the detector surface were etched for 5 h in 6N KOH + 20 % ethyl alcohol. The heights of more than 4000 etch-pit cones on each Makrofol detector were measured using an optical microscope with a 400x magnification, coupled to a CCD camera and a video monitor. We measured, with an accuracy of ± 1 μm, the cone heights $\geq$ 75 μm for the detector exposed at 90°; for the detector exposed at 45° the cone heights $\geq$ 4 μm are easily measurable. The corresponding distributions are shown in Figure 3(a,b); for normal incidence (a) each peak is well separated from the others, and a charge can be assigned to every peak. Notice the Z=83 peak which corresponds to a charge pick-up reaction. For 45° incidence (b) we can distinguish fragment peaks with $Z \geq 59$.

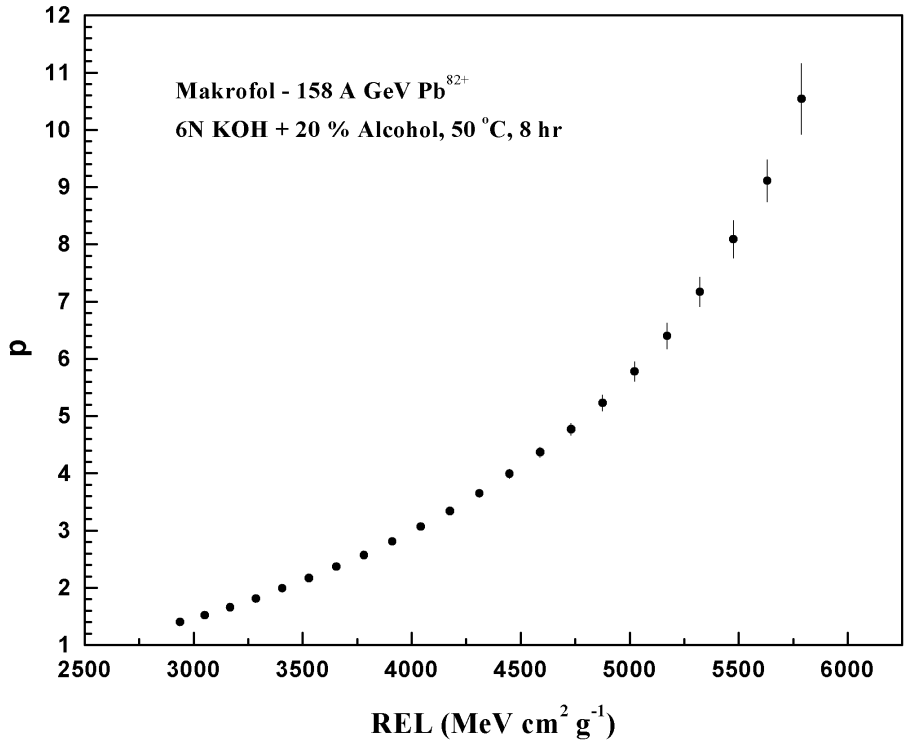

Figure 4.  Reduced etch rate $p = v_T/v_B$ vs REL for the Makrofol detector exposed to a relativistic lead beam.

For each nuclear fragment detected in the NTDs exposed at normal incidence we computed the Restricted Energy Loss (REL) and the reduced etch rate $p = v_T/v_B$ using the formula (Giacomelli et al., 1998).

$$p = \frac{1 + (D/2v_B t)^2}{1 - (D/2v_B t)^2} \quad (1)$$

where 'D' is the diameter of the track base area, '$v_B$' is the detector bulk etch rate and 't' is the etching time.

The bulk etching rate was determined by measuring the mean thickness difference before and after etching with an accuracy of ± 1 μm (Dwivedi and Mukherji, 1979).

The reduced etch rate 'p' versus 'REL' for the Makrofol NTD is plotted in Figure 4. The detection threshold is at REL ~ 2500 MeV cm$^2$ g$^{-1}$, corresponding to a relativistic nuclear fragment with charge $Z \geq 50$.

## 4. Conclusions

The results of the new chemical etching process for Makrofol NTDs with the addition of ethyl alcohol in the solution improves the post etched detector surface, removes background tracks and enhances the sharpness of the tracks.

Different methods were used to measure the cone heights and base areas of the detected nuclei (Pb-ions + their fragments). For the first time one observed well separated peaks for the lead fragments. Extensive calibrations of the Makrofol detectors were made. Using relativistic Pb-ions, we established the detection threshold of Makrofol to be $Z/\beta \sim 50$.

The new etching conditions allowed the clear observation of single peaks both in base area (Fig. 2) as well as in cone height distributions (Fig 3).

**Acknowledgements**

We thank the CERN SPS staff for the Pb exposure. We acknowledge many colleagues for their cooperation and technical advices. We gratefully acknowledge the contribution of our technical staff: E. Bottazzi, L. Degli Esposti, R. Giacomelli, G. Grandi and C. Valieri. We thank INFN and ICTP for providing fellowships and grants to non-Italian citizens.